# Can Computer Algebra be Liberated from its Algebraic Yoke ?
## An Algorithmic Approach to Functional Analysis from a Functional Approach to Computing

Rémi Barrère
Université de Franche-Comté, ENSMM
26, chemin de l'Epitaphe, F-25000 Besançon, France
Fax : 33(0)381 809 870
E-mail : rbarrere@ens2m.fr

## Abstract

So far, the scope of computer algebra has been needlessly restricted to exact algebraic methods. Its possible extension to approximate analytical methods is discussed. The entangled roles of functional analysis and symbolic programming, especially the functional and transformational paradigms, are put forward. In the future, algebraic algorithms could constitute the core of extended symbolic manipulation systems including primitives for symbolic approximations.

## 1 Introduction

In recent decades, the extensive development of numerical techniques entailed a lower use of traditional approximate analytical methods, such as series expansions or functional iteration. Admittedly, computer algebra is currently reviving the analytical approach, and active researches in that area bring about new ideas and let expect future progresses (see for instance the Journal of Symbolic Computation or the proceedings of the ISSAC conferences). The underlying algorithms yield symbolic expressions that are generally more compact, easier to manipulate and more meaningful than numerical data.

However, computer algebra techniques are more or less restricted to exact solutions, so they often fail in the case of physical or engineering problems. Indeed, mathematical modelling commonly leads to equations that have no known closed-form solution. In that case the general trend consists in resorting to numerical techniques. Unfortunately, these have drawbacks such as the difficulty to link different computations or to interpret purely numerical results: the lack of symbolic representation prevents the user from identifying patterns among models, hence from finding relevant generic models.

The purpose of the paper is to show that symbolic manipulations are still worthwhile when associated with functional methods. Here, "functional" refers to the cooperation of function theoretic methods and the functional programming paradigm. Indeed, most function space methods are "constructive" in the sense that they basically generate symbolic approximations. Moreover, the resulting algorithms turn out to be well adapted to the functional paradigm. So functional programming can be regarded as the natural computational counterpart of functional analysis.

## 2 Functional and rule-based programming

The algorithms described below were drafted in Mathematica, which consistently combines a computer algebra system with a mutiparadigm programming language built on top of a rule-based core. Its sophisticated pattern-matching capabilities facilitate the writing of concise symbolic programs, the syntax of which mimics that of mathematics [31] (see also the critical reviews by Fateman [14, 15] and Maeder's presentations [19, 21]).

Actually, Mathematica results from a symbiosis of the functional and rule-based (transformational) paradigms: the user defines functions which are internally processed as transformation rules by a rewriting mechanism. As a functional language, Mathematica can be thought of as a descendant of Lisp; as a transformational one, it can also be regarded as a descendant of Snobol with a noticeable influence of Prolog.

Born in the late fifties with Lisp in a context of artificial intelligence, the functional paradigm really took off in the eighties thanks to novel design techniques such as graph rewriting [26]. With a functional language, a computation is processed as a function evaluation. The theory of recursive functions and the $\lambda$-calculus constitute the mathematical foundation of the functional paradigm.

Born in the mid sixties with Snobol, the transformational paradigm remained more or less hidden behind the scenes until it was acknowledged as a genuine paradigm on the occasion of researches in programming paradigms [27]. Rule-based languages emphasize the transformation of symbolic expressions according to their form, so a computation amounts to the restructuring of an expression by term rewriting. The Post-Markov theory of algorithms constitute the mathematical counterpart of this paradigm.

These languages enable an abstract programming style thanks to high level constructs such as patterns, rewrite rules or higher-order functions that avoid classical control structures. They proved to be a sound basis for computer algebra systems (e.g., Reduce, Macsyma), so the primitives of the language can be combined with the mathematical operators, thus leading to a concise programming style where computing and programming tend to blend into a single activity [5].

By combining the functional and transformational paradigms, Mathematica facilitates the recourse to denotational semantics either as a tool for program modelling or more generally for translating relational algebraic formulations into transformational algorithmic ones (Knuth-Bendix algorithm or its variants).

These considerations are far from superfluous. Computer scientists have become aware of the role of "programming languages as thought models" [29], which is of particular interest in scientific computing. They now take into account that "the language in which a programmer thinks a problem will be solved will color and alter, in a basic and fundamental way, the fashion in which that programmer will develop an algorithm" [11]. In particular, thanks to higher-order functions, Mathematica facilitate the implementation of functional methods, which lend themselves well to symbolic programming.

## 3 Role of symbolic approximations

Symbolic approximations combine the advantages of symbolic computations and approximate methods. As opposed to numerical techniques, they enable the manipulation of literal expressions with symbolic parameters, thus entailing a better understanding, hence a better mastery of the underlying physical systems.

Actually, physical or engineering applications do not necessarily require exact solutions nor accurate approximations. Indeed, most (if not all) models are based on assumptions, simplifications or approximations, so it does not necessarily make sense to seek exact solutions nor very precise approximations.

Moreover, exact solutions occasionally have such a complexity that approximations may be more practical. Sometimes, even rough representations may give more insight than intricate exact solutions or bulky numerical results. Then, approximating can be a way of simplifying.

A few fundamental methods from functional analysis (e.g., fixed points, perturbation expansions or variational formulations) constitute the theoretical background for this processes. The present investigation focusses on some of their algorithmic counterparts, namely successive approximations, Newton's method, perturbation series and the Galerkin procedure.

A comment might be worth stressing here in order to dismiss a common misinterpretation: function theoretic methods are inherently symbolic (precisely because they apply to function spaces), although they are usually thought of as numerical. In particular, they require algebraic facilities, hence rely on the algebraic core of computer algebra.

These methods have been used long, even before the emergence of computing, but they are tackled here as generic tools for the programming of symbolic approximate methods. Papers referring to this approach are still scattered about the scientific literature, which testifies its structuring power in scientific computing has not been fully recognized yet.

## 4 Fixed points and successive approximations

**Principle and programming of the method**

Although it is often restricted to existence and uniqueness demonstrations [22], the method of successive approximations also turns out to be valuable for computing approximate (occasionally exact) symbolic solutions. The method is valid in (complete) metric spaces, so it applies to numerical as well as functional or even geometrical problems. Its statement can be briefly recalled; as long as f is a contraction mapping (f satisfies a Lipschitz condition with k<1) for some distance:

$$d[f(x),f(y)] \leq k\, d(x,y) , \qquad (1)$$

we can solve the equation f(x)=x, i.e., we can find the (unique) fixed point of f, by computing the limit:

$$x_\infty = \lim_{n \to \infty} f^{on}(x) \qquad (2)$$

where $f^{on}$ denotes the $n^{th}$ iterate. Moreover, provided some initial approximation $x_o$ is in the contraction domain of f, any iterate

$$x_n = f^{on}(x_o)$$

can be considered as an approximation to the fixed point $x_\infty$. Such approximants are sometimes called continued-functions or iterated functions. Above all, functional languages have an operator ("Nest" in Mathematica) that computes the $n^{th}$ iterate, so the program comes down to using a primitive of the language. When f is an operator in a function space, its algorithmic counterpart is a higher-order function.

The example below shows the first approximation to the solution to the equation u′+(1+φ$^2$)u/2=φ, with u(0)=0 and φ=a sin(wt), drawn from a model of an electric motor. The second approximation enabled the identification of a periodic steady-state behaviour, which was validated by comparison with a numerical approximation.

```
x[t_]:= a Sin[w t]
operator[u_][t_]:= Integrate[
   Exp[-(t-τ)/2]*(x[τ]-u[τ]*x[τ]^2/2),
   {τ,0,t}
]
Nest[operator[#][t]&, 0 &[t], 1]
```

$$\frac{4aw}{E^{t/2}(1 + 4w^2)} - \frac{2a\ (2\ w\ Cos[tw] - Sin[tw])^2}{1 + 4w^2}$$

Fixed point equations can be written in different ways (e.g., f(x)=x <=> x=f$^{-1}$(x) ), which is of practical importance for finding a form where f is a contraction with the smallest possible value of k. For instance, it can be necessary to turn a differential equation into an integral equation.

**Convergence considerations**

The major restriction regards the Lipschitz condition, since the convergence is limited to the contraction domain of f. When it is assured, the method leads to a linear convergence, i.e., $d(x_{n+1},x_\infty) \le k\, d(x_n,x_\infty)$. Nevertheless, the non-linear Aitken-Shanks transformation improves the rate of convergence [9].

$$\begin{aligned} a_n = a(x_n) &= \frac{x_{n+2}\ x_n - x_{n+1}^2}{x_{n+2} + x_n - 2x_{n+1}} \\ &= x_{n+2} + \frac{1}{\dfrac{1}{x_{n+2} - x_{n+1}} - \dfrac{1}{x_{n+1} - x_n}} \end{aligned} \quad (3)$$

It (the second form) is generally used in numerical contexts, although there is no impediment for its use in a symbolic context. In the case of successive approximations, its efficiency results from the fact that it extrapolates the (asymptotic) geometric behaviour of the sequence. In that case, the algorithm (called Steffesen's method) leads to the following expression, the programming of which is straightforward in a functional style.

$$\begin{aligned} s_n = s(x_n) &= \frac{f^{o2}(x_n)\ x_n - f(x_n)^2}{f^{o2}(x_n) + x_n - 2f(x_n)} \\ &= \frac{f^{o(n+2)}(x_0)\ f^{on}(x_0) - f^{o(n+1)}(x_0)^2}{f^{o(n+2)}(x_0) + f^{on}(x_0) - f^{o(n+1)}(x_0)} \end{aligned} \quad (4)$$

## 5 The Newton method

**Principle of the method**

Although it is commonly thought ofas a numerical technique, Newton's method applies to Banach spaces, so it basically solves functional equations in a symbolic way. This was mentioned by Kantorovich [17] and to some extent used by Bellman [7] who called it quasilinearization, for it relies on solving a sequence of linear equations. Since then, the nascent idea seems to have been overwhelmed by the numerical tidal wave. It should not be likened to the particular use of Newton's method for series computations [18].

Besides, optimization problems are connected with it since finding an extremum to g amounts to seeking a zero to f=grad g (hence the idea of approximate symbolic optimization). The following example is worth mentioning for it is often used as a test for numerical algorithms: the (exact) extremum (1,1) of the Rosenbrock function f(x,y) = (x-1)$^2$+p (x$^2$-y)$^2$ turns out to be reached in two steps, from the initial approximation (x,y)=(0,0).

```
Rosenbrock[p_,x_,y_]:=(x-1)^2+p(x^2-y)^2
equations={D[Rosenbrock[p,x,y],x]==0,
           D[Rosenbrock[p,x,y],y]==0}
NewtonASolve[equations, {{x,y}, {0,0}, 2},
           Simplify->True]
{{y -> 1, x -> 1}}
```

Here is the schematic statement of Newton's method: as long as f has an invertible Fréchet derivative grad f, and the initial approximation $x_0$ is sufficiently near the solution (which can be more precisely stated for specific problems in connection with convexity properties), a sequence of approximate solutions to the equation f(x)=0 can be computed by means of the recurrence formula:

$$x_{n+1} = x_n - \left[\text{grad } f(x_n)\right]^{-1}[f(x_n)] \quad (5)$$

In other words, $x_\infty$ is the fixed point of x-[grad f (x)]$^{-1}$[f(x)]. Since it is a functional iteration, the procedure can also be implemented in a functional language by means of an iteration operator, but it requires further programming for computing the Fréchet derivative. In the case of differential or integral equations, the recurrence formula is rather written in the following form, which actually designates a sequence of linear equations to be solved:

$$\text{grad } f(x_n)\left[x_{n+1} - x_n\right] + f(x_n) = 0 \quad (6)$$

The Prager model for the deflection of a beam [7] gives an example of a non-linear Sturm problem : u''+(a u')$^2$+1=0,  u(0)=u(1)=0. Below, the first approximation is computed (with $u_0(x)$=0 as initial approximation) and is shown together with the exact solution (in solid line).

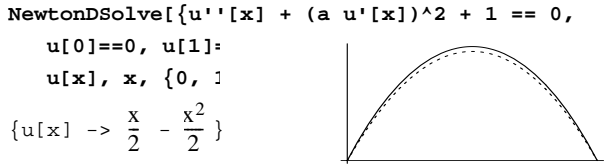

```
NewtonDSolve[{u''[x] + (a u'[x])^2 + 1 == 0,
   u[0]==0, u[1]=
   u[x], x, {0, 1
{u[x] -> x/2 - x^2/2 }
```

**Figure 1**: Prager's model for the deflection of a beam.

**Programming the Fréchet derivative**

According to the circumstances, the Fréchet derivative (or tangent linear map) can be expressed by means of the ordinary derivative, the jacobian matrix or the gradient operator. In this last case, it would be unreasonable to aim at an algorithm computing any Fréchet derivative. But, a quite general program can be designed for processing the functional expressions commonly encountered when solving differential or integral equations. It is based on a few usual derivatives and the following formulas [22], where FD[f] denotes the Fréchet (or functional) derivative of f with respect to φ :

$$FD[f(\varphi(x), \varphi'(x), \ldots, \varphi^{(n)}(x), x)][h(x)]$$
$$= \sum_{i=1}^{n+1} \partial_i f(\varphi(x), \varphi'(x), \ldots, \varphi^{(n)}(x), x) h^{(i-1)}(x)$$

$$FD\left[\int_\Omega f(\varphi(x), \varphi'(x), \ldots, \varphi^{(n)}(x), x) dx\right][h(x)] \quad (7)$$
$$= \int_\Omega FD[f(\varphi(x), \varphi'(x), \ldots, \varphi^{(n)}(x), x)][h(x)] dx$$

Pattern-matching and transformation rules facilitate the programming of such formulas. Here is a short extract of the program.

```
FDerivative[
   Integrate[expr_,{x_,a_,b_}],u_[x_]
][h_]:=
   Integrate[
      FDerivative[expr,u[x]][h],
      {x,a,b}
   ]/;FreeQ[a,x]&&FreeQ[b,x]
```

A subsidiary application of the Fréchet derivative is worth mentioning in this context. It regards the computation of variational formulations associated with potential operators in the case of conservative systems [25].

```
FDerivative[
   Integrate[
      1/2 u'[x]^2 - f[x] u[x],
      {x,a,b}
   ], u[x]][v[x]]
Integrate[u'[x] v'[x] - f[x] v[x],{x,a,b}]
```

The example concerns the common case of a quadratic functional, which leads to the usual second degree form.

**Discussion and future directions**

The quadratic convergence, namely $\|x_{n+1}-x_\infty\| \le k \|x_n-x_\infty\|^2$, is an advantage of Newton's method; this generally compensates its greater computational complexity. However, Fateman [13] produced examples for which the quadratically convergent algorithm is only marginally faster and sometimes slower than a linear one. Nevertheless, in most circumstances, quasilinearization is known to provide with a wider interval of convergence than successive approximations. A drawback may be the difficulty for computing the inverse of the Fréchet derivative. This problem especially arises with (non-linear) differential equations, for the Fréchet derivative generally leads to (linear) differential equations with variable coefficients. Nevertheless in that case, approximations to grad f may lead to satisfactory results [7], which means for instance, the possibility of combining the Newton method with the Galerkin one (section 7) for solving the associated linear problems.

## 6 Perturbation expansions

**Principle of the method and functional formulation**

Only regular perturbations are discussed here. Perturbation series arise from equations of the form:
$$f(x, \varepsilon) = 0 \quad (8)$$
where f is an operator between Banach spaces, which means the algorithm is valid whatever the precise nature of the equation. It is assumed $f(x,0)=0$ can be solved and the parameter ε is supposed to be small, so $f(x, \varepsilon) =0$ can be viewed as a perturbed model of $f(x,0)=0$; likewise $f(x,0)=0$ can be viewed as an approximate model of $f(x, \varepsilon)=0$.

The perturbation procedure is related to the implicit function theorem [16]: if $f(x, \varepsilon)$ is a $C^k$ mapping between Banach spaces, and there is an $x_0$ satisfying $f(x_0,0)=0$ such that the partial (Fréchet) derivative $\partial_x f(x_0,0)$ is an invertible linear map, then there is a unique solution to the implicit equation $f(x, \varepsilon)=0$ in the neighborhood of ε=0, given by $x=\varphi(\varepsilon)$, where φ is a $C^k$ mapping.

The procedure for regular perturbations [6] consists in seeking a formal solution in the form of a power series in epsilon:
$$x=g(\varepsilon)=\sum \varepsilon^k x_k \quad (9)$$
where the $x_k$ are the unknowns. The method of indeterminate coefficients gives rise to an infinite sequence of equations that can be solved recursively:
$$f_k(x_{i,i\le k})=0 \quad (10)$$
In practice, a truncated series is computed by solving a finite system. Obtaining a functional iterative formulation is a bit tricky, but this can be achieved by way of the accumulation operator "Fold" in Mathematica.

As an illustrative example, here is a perturbative approximation to the Bernoulli differential equation $u'+k u+e u^2=0, u(0)=1$.

```
theSeries=PerturbationADSolve[{
   D[u[t],t]+k u[t]+e u[t]^2==0,
   u[0]==1
   },u[t],t,{e,1}]
```

$\{u[t] \to e^{-kt} + \frac{e}{k}(e^{-2kt} - e^{-kt})\}$

**Approximate resummation**

Perturbation expansions tend to generate slowly convergent series or even divergent ones. It has been a common practice to improve the convergence by means of Padé approximants, which in fact constitute a resummation procedure [9]. The idea of Padé summation is to replace a truncated power series by a rational function whose first terms in the series expansion match the given series. Its advantage comes from the fact that its computation involves only algebraic manipulations. Although Padé approximants have been used essentially from a numerical point of view, they can be implemented in a symbolic way. Moreover, they are known to work well, even beyond their proven range of applicability [1].

In the previous example, the Padé approximant $P_0/Q_1$ restores the exact solution from the second approximation.

```
Needs["Calculus`Pade`"]
Pade[u[t]/.theSeries,{e,0,0,1}]//Simplify
```

$\frac{k}{-e + (e+k)E^{kt}}$

Padé approximants are at the heart of a more general resummation method suggested by Bergeron and Plouffe [10] in the context of combinatorics. An extension of their heuristic procedure to perturbation series was suggested in [3].

**Future work**

The perturbation method often leads to non uniform (singular) perturbations, and non-linear equations may generate bifurcation phenomena. Appropriate computational techniques have been settled to overcome those difficulties [24, 28]. Their implementation as generic operators could be investigated. More generally, a full implementation of the perturbation method would facilitate the symbolic treatment of homogeneization or boundary layer problems.

# 7 Weak solutions and the Galerkin procedure

**Principle of the method**

The Galerkin method regards boundary-value problems, which typically arise from continuum physics models; here, we will limit ourselves to linear equations. Most of these models can be placed into one of the three following classes: elliptic (or Laplace) equations for static phenomena, spectral (or Helmholtz) problems for stationary phenomena, time evolution (parabolic or hyperbolic) equations in the case of dynamic behaviour (diffusion or wave propagation). This classifying turns out to be useful from the algorithmic point of view. The method is closely related to weak (or variational) formulations, which provide with both theoretical results and approximation techniques [8, 30]. Schematically, starting with the initial differential formulation with its boundary conditions :

$$Au=f \qquad (11)$$

(respectively : $Au=\lambda u$, or $Au=D_t u$), the approximation is obtained in two stages. First, a weak formulation leads to an integral expression based on a bilinear operator, i.e., a functional inner product :

$$\forall v, <Au,v>=<f,v>, \text{ where } : <u,v>=\int_\Omega u\, v\, d\Omega \qquad (12)$$

(respectively : $<Au,v>=\lambda<u,v>$, or $<Au,v>=D_t<u,v>$). Then an approximate solution $u_n$ is searched in the form of a linear combination of n (previously chosen) basis functions :

$$u_n=\sum c_i w_i \qquad (13)$$

In other words, $u_n$ is the projection of the exact solution onto the finite dimensional subspace spanned by the system of functions $w_i$. The test functions v and the basis functions $w_i$ are supposed to belong to a convenient Sobolev space. This leads to a finite set of algebraic (respectively proper value, or differential) equations.

$$\sum_1 c_i \langle A w_i, w_j \rangle = \langle f, w_j \rangle, j \in \{1, n\} \qquad (14)$$

respectively :

$$\sum_1 c_i \langle A w_i, w_j \rangle = \lambda \sum_1 c_i \langle w_i, w_j \rangle$$

$$\sum_1 c_i(t) \langle A w_i, w_j \rangle = D_t \sum_1 c_i(t) \langle w_i, w_j \rangle$$

Piecewise linear or piecewise polynomial approximations can be obtained by the finite element technique; this is a common numerical approach. Nevertheless, other kinds of basis functions can be chosen, according to the shape of the domain, orthogonality considerations or possibly experimental data. On the one hand, purely numerical solutions are obtained, by solving large systems; no symbolic parameter is available, but intricate boundaries are possible. On the other hand, approximate symbolic solutions are reachable, by solving small systems; analytical expressions with symbolic parameters can be manipulated, but intricate boundaries are difficult to handle.

**Functional formulation**

The procedure is entirely determined by inner products and a truncated system of functions, so it lends itself well to the functional style. In fact, according to whether the problem is elliptic, a spectral problem or an evolution equation, three generic operators are required, the arguments of which are the inner products and the basis [2]. Here is the first one (the others being similar), where "Outer" (generalized tensor product) generates the left-hand matrix, whereas "Map" generates the right-hand vector. Less generic programs are given in [8] or [23].

```
GalerkinSolve[a_,l_,basis_List]:=
  LinearSolve[
    Outer[a,basis, basis],
    Map[l, basis]
  ].basis
```

However, this simplicity has a drawback: constructing the inner products, which characterize each particular problem, is up to the user and may be a more or less difficult task. The worthwhile counterpart is the possibility of choosing a purely symbolic, a purely numerical or a mixed symbolic-numerical solution. Transferring this choice on the definitions of the inner products makes the procedure "GalerkinSolve" fully generic.

In the exemple, the finite element technique (approximation with 3 hat functions without interpolation) was qualitatively compared with the symbolic Galerkin method (3 trigonometric functions) for the 1-D Laplace equation: $-u''=e^x \sin 5x$ with homogeneous boudary conditions.

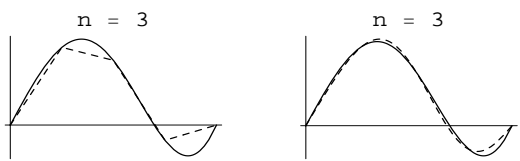

**Figure 2:** finite elements vs. symbolic Galerkin.

**Possible developments**

The Galerkin procedure constitutes the springboard for a mixed symbolic-numerical treatment of continuum physics equations. As an illustration of this expanding approach, Dasgupta [12] resorted to the symbolic capabilities of Mathematica to generate shape functions for concave quadrilateral elements by means of Padé approximants.

In the case of nonlinear problems, the finite element technique is quite usually associated with a numerical version of Newton's method. Similarly, the Galerkin procedure could be associated with the symbolic version of Newton's method or with the perturbative approach.

## 8 Discussion

**Symbolic approximations**

The aforementioned operators for symbolic approximations can be used like built-in ones. For instance, they can be nested or combined with each other, so that intermediate results can be "piped" from (or to) each other. Actually, they extend the capabilities of the software environment, where most computations tend to be entirely processed.

In practice, this kind of computation requires a permanent control from the user, which is a characteristic of most computer algebra manipulations. In particular, the importance of a judicious choice of some parameters (e.g., initial approximations or basis functions) must be stressed, so an intuitive knowledge of the solutions can be of great service.

These methods are (or can be made) "adaptive" in the sense that the approximations can be refined step by step; so the user can choose any intermediate between rough cheap approximations, and more precise but more expensive ones. If need be, these can be implemented by means of a stream [5, section 3.2.5].

As opposed to numerical computations, symbolic manipulations pave the way for the discovery of relevant patterns among the models and their solutions, which enable in turn the emergence of novel mathematical abstractions.

However, these methods still rely on computer algebra capabilities, the current limitations of which sometimes hinder the whole computation. In particular, like algebraic manipulations, analytical ones tend to generate intricate expressions; in such cases, only low-order approximations are practically reachable.

**Symbolic programming**

Functional languages enable the manipulation of higher-order functions which are particularly useful for implementing function theoretic algorithms. Also, pattern-matching facilitates the design of generic operators that apply to classes of arguments (equations in this context). This actually means that more or less general mathematical methods can be translated into generic operators. Broadly speaking, symbolic languages facilitate the expression of scientific knowledge and the programming of mathematical abstractions.

For historical reasons, programs written in symbolic languages have obtained a reputation for lack of performance. Part of this results from the high-level of abstraction that is available in such languages and from the background operations that they require. However, in the eighties, research in functional languages resulted in new design techniques leading to better performances [26]. Besides, although procedural programming is known to be more efficient for numerical computations, functional programming often turns out to be more efficient for symbolic ones [20].

Also, the spreading of software environments that combine a computer algebra system with a programming language arouses new activities in mathematical modelling and scientific computing where the development time reaches the same order of magnitude as the time of use; so it can no longer be neglected, and benchmarks should now take this into account. Furthermore, under the influence of computer algebra and symbolic programming, scientists become aware of the role of programming languages as a medium of scientific knowledge, so in some circumstances, the form of a program may take priority over its efficiency.

Finally, the referential transparency of (purely) functional programs facilitates their mathematical description. Actually, the programs and their descriptions tend to become two complementary aspects of a single formulation: a transformational version and a relational one. Moreover, the emergence of very high-level languages that implement the main three theories of computation, thus enabling the direct expression of mathematical knowledge, tends to extenuate the traditional distinction between algorithms and programs [4].

## 9  Conclusion

When equations have no known closed-form solution, symbolic approximations may constitute an alternative to numerical techniques, thanks to the symbiosis of the expressive power of symbolic languages and the computational power of function space methods.

On the one hand, these turn out to be a uniform conceptual approach to the study of functional equations: they provide not only potent theoretical tools (existence, uniqueness and convergence theorems), but also efficient symbolic approximation procedures. In that sense, these are constructive methods.

On the other hand, the resulting algorithms lend themselves well to the functional and transformational programming paradigms. They are quite easily implemented as generic operators in a computer algebra and symbolic programming compound environment.

To sum up, not only functional analysis is the natural theoretical background for symbolic approximation theory, but also functional programming turns out to be its natural computational counterpart. Together, they pave the way for an algorithmic approach to functional analysis supported by a functional approach to computing. In the future, the integration of symbolic approximation algorithms into computer algebra systems might contribute to extending the restrictive connotation of computer algebra to the wider notion of analytical manipulation.

A set of experimental Mathematica packages and notebooks, including the code and a documentation is freely available (http://macmaths.ens2m.fr/research/pages/symbolic.html).